\documentclass[pra,twocolumn,nofootinbib,floatfix]{revtex4-1} 
\usepackage[usenames]{color}
\usepackage{hyperref}

\usepackage{bbold}

\usepackage{hhline}

\usepackage[utf8]{inputenc}
\usepackage[T1]{fontenc}
\usepackage[pdftex]{graphicx}

\usepackage{amsmath, amsxtra, amssymb, mathtools, isomath, xspace} 
\renewcommand{\vec}{\vectorsym}
\usepackage{siunitx}

\usepackage[british]{babel}

\usepackage{booktabs}
\usepackage{multirow}

\newcommand{\mycite}[1]{\cite{#1}}

\newcommand{\ket}[1]{\ensuremath{\vert #1 \rangle}\xspace}%
\newcommand{\bra}[1]{\ensuremath{\langle #1 \vert}\xspace}%
\newcommand{\avg}[1]{\ensuremath{\langle #1 \rangle}\xspace}%
\newcommand{\Er}{\ensuremath{E_\text{r}}\xspace}%

\newcommand{\Rb}{\ensuremath{^{87}\text{Rb}}\xspace}%
\newcommand{\Rbk}{\ensuremath{R_{\text b}}\xspace}%
\newcommand{\Nat}{\ensuremath{{N_{\text{at}}}}\xspace}%
\newcommand{\Ne}{\ensuremath{{N_{\text{e}}}}\xspace}%
\newcommand{\alat}{\ensuremath{a_{\text{lat}}}\xspace}%

\newcommand{\gtwo}{\ensuremath{g^{(2)}}\xspace}%

\long\def\symbolfootnote[#1]#2{\begingroup%
\def\thefootnote{\fnsymbol{footnote}}\footnotetext[#1]{#2}\endgroup}

\begin{document}

\title{Observation of mesoscopic crystalline structures in a two-dimensional Rydberg gas}


\author{Peter~Schau\ss$^{1,*}$}%
\author{Marc Cheneau$^{1}$}%
\author{Manuel Endres$^{1}$}%
\author{Takeshi Fukuhara$^{1}$}%
\author{Sebastian Hild$^{1}$}%
\author{Ahmed Omran$^{1}$}%
\author{Thomas Pohl$^{2}$}%
\author{Christian Gross$^{1}$}%
\author{Stefan Kuhr$^{1,3}$}%
\author{Immanuel Bloch$^{1,4}$}%

\date{29 August 2012}

\affiliation{$^1$Max-Planck-Institut f\"{u}r Quantenoptik, 85748 Garching, Germany}
\affiliation{$^2$Max-Planck-Institut f\"ur Physik komplexer Systeme, 01187 Dresden, Germany}%
\affiliation{$^3$University of Strathclyde, Department of Physics, SUPA, Glasgow G4 0NG, UK}%
\affiliation{$^4$Ludwig-Maximilians-Universit\"{a}t, Fakult\"{a}t f\"{u}r Physik, 80799 M\"{u}nchen, Germany}%

\begin{abstract}
The ability to control and tune interactions in ultracold atomic gases has paved the way towards the realization of new phases of matter. Whereas experiments have so far achieved a high degree of control over short-ranged interactions, the realization of long-range interactions would open up a whole new realm of many-body physics and has become a central focus of research. Rydberg atoms are very well-suited to achieve this goal, as the van der Waals forces between them are many orders of magnitude larger than for ground state atoms \mycite{Saffman:2010}. Consequently, the mere laser excitation of ultracold gases can cause strongly correlated many-body states to emerge directly when atoms are transferred to Rydberg states. A key example are quantum crystals, composed of coherent superpositions of different spatially ordered configurations of collective excitations \mycite{Weimer:2008, Pohl2010a, Schachenmayer2010, Garttner2012}. Here we report on the direct measurement of strong correlations in a laser excited two-dimensional atomic Mott insulator \mycite{Bloch:2008c} using high-resolution, in-situ Rydberg atom imaging. The observations reveal the emergence of spatially ordered excitation patterns in the high-density components of the prepared many-body state. They have random orientation, but well defined geometry, forming mesoscopic crystals of collective excitations delocalised throughout the gas. Our experiment demonstrates the potential of Rydberg gases to realise exotic phases of matter, thereby laying the basis for quantum simulations of long-range interacting quantum magnets.
\end{abstract}

  \symbolfootnote[1]{Electronic address: {\bf peter.schauss@mpq.mpg.de}}
  \maketitle%
The strongly enhanced interaction between Rydberg atoms makes them unique building blocks for a variety of applications ranging from quantum optics and quantum information processing \mycite{Jaksch2000, Lukin2001,Saffman:2010} to engineering of exotic quantum many-body phases \mycite{Pupillo:2010, Henkel2010a, Cinti2010}. For the latter purpose, two main ideas have been explored theoretically. On the one hand, the weak admixing of a Rydberg state to the atomic ground state using off-resonant laser coupling was suggested as a way to benefit from the long-range interactions without persistent population in the Rydberg state \mycite{Pupillo:2010, Henkel2010a, Honer:2010}. On the other hand, direct laser excitation leads to the formation of a gas of Rydberg excitations, also called Rydberg gas. This strongly correlated system \mycite{Robicheaux2005} can exhibit highly non-classical states characterized by the coherent superposition of ordered structures in the spatial distribution of the Rydberg excitations \mycite{Weimer:2008, Pohl2010a, Schachenmayer2010, VanBijnen2011, Garttner2012, Lesanovsky2011,Hoening2012}. Here the excitation dynamics proceeds on a timescale of a few microseconds, on which the atoms can be considered frozen in space, representing strongly interacting effective spins. At the heart of the formation of such correlated states lies the dipole blockade effect \mycite{Jaksch2000, Lukin2001, Saffman:2010} that prevents simultaneous Rydberg excitation of two close-by atoms \mycite{Tong2004, Urban2009, Gaetan2009, Schwarzkopf2011, Dudin2012a}. Recent experiments using two trapped atoms have shown how this blockade effect can be used to implement fast two-qubit quantum gates \mycite{Wilk:2010,Isenhower2010}. In larger ultracold atomic ensembles, the coherence of the collective excitation was demonstrated \mycite{Raitzsch2008, Reetz-Lamour2008, Dudin2012b} and evidence for strong correlations could be found by observing universal scaling laws for the number of excited Rydberg atoms \mycite{Low2009, Viteau2011a}. However, direct measurements of spatial ordering have remained an outstanding challenge. Important steps in this direction were recently explored using a field-ion-microscope \mycite{Schwarzkopf2011}, allowing for the measurement of the blockade radius in a three-dimensional Rydberg gas. Recent theoretical works, on the other hand, have proposed detection schemes with potential resolution below the blockade radius, based on conditional Raman transfer \mycite{Olmos2011} or electromagnetically induced transparency \mycite{Gunter2011}.

  Here, we demonstrate an alternative approach that permits direct imaging of spatial excitation patterns, and precise measurements of correlation functions. This allows to probe the underlying spatially ordered constituents of the excited many-body state, revealing  crystalline excitation patterns of its high-density components. Two key advancements form the basis of our observations. First, a two-dimensional atomic Mott insulator provides a dense and well-ordered initial system that maximises coherence times during the excitation dynamics. Second, we developed an all-optical technique to image individual Rydberg atoms in-situ with high spatial and temporal resolution.
 
  

  \begin{figure}
    \centering 
     \includegraphics{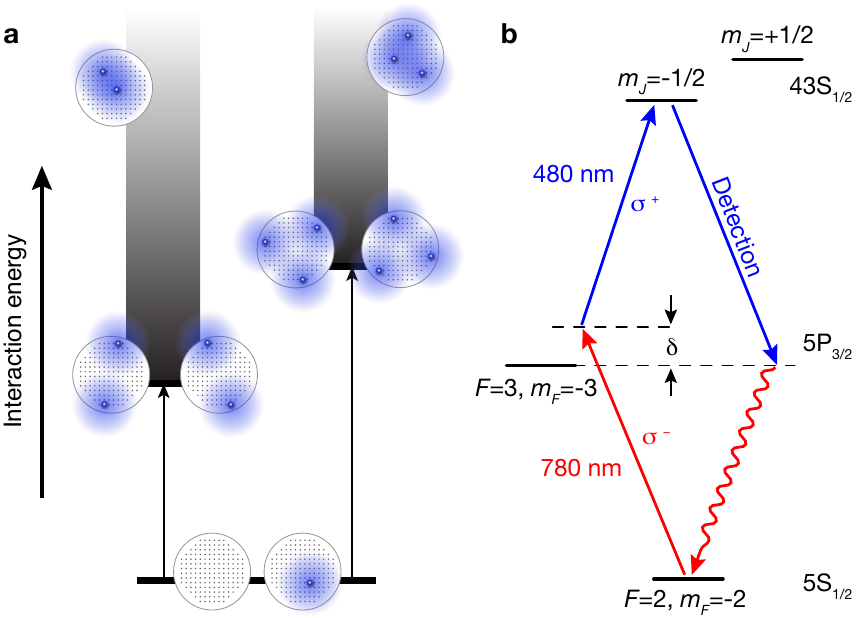}
    \caption{{\bf Schematics of the many-body excitation.}  \textbf{a}, Energy spectrum in the
      absence of optical driving. States with more than one excitation form a broad energy band
      (gray shading) above the degenerate manifold comprising the ground state and all singly
      excited states. For each excitation number \mbox{$\Ne>1$}, the states with lowest energy
      correspond to spatially ordered configurations, which maximise the distance between the
      Rydberg excitations. The minimal interaction energy (black arrows) is determined by the finite
      system size and increases with $\Ne$. Possible spatial configurations of the excitations (blue
      dots) in the initial Mott-insulating state (black dots) are shown schematically as circular
      insets next to their respective interaction energy. The blockade radius is depicted by the
      blue shaded disc around the excitation. \textbf{b}, Simplified level scheme of
      $^{87}\text{Rb}$ showing the transitions used for the Rydberg excitation and detection.}
    \label{fig:1}
  \end{figure}
  

  The physical system considered here is a two-dimensional gas of alkali atoms trapped in a
  rotationally invariant harmonic confinement potential and pinned in a square optical lattice. The
  gas was prepared deep in the Mott-insulating phase, ensuring uniform filling with one atom per
  site within a disk of radius $R \simeq \sqrt{\Nat a_{\text{lat}}^2/\pi}$, where \Nat is the total
  number of atoms and $\alat$ the lattice spacing. The atoms were initially in their electronic
  ground state, \ket{g}, and then resonantly coupled to a Rydberg state, \ket{e}. In the interaction
  picture, the internal dynamics of the atoms is governed by the many-body Hamiltonian:
  \begin{equation}
    \label{eq:H}
    \hat{H} = \frac{\hbar\Omega}{2} \sum_{\vec i} \left( \hat\sigma^{(\vec i)}_{eg} + \hat\sigma^{(\vec i)}_{ge} \right)
    + \sum_{\vec i \neq \vec j} \frac{V_{\vec i \vec j}}{2} \hat\sigma^{(\vec i)}_{ee} \hat\sigma^{(\vec j)}_{ee} \; .
  \end{equation}
  Here, the vectors $\vec i = (i_{x}, i_{y})$ label the lattice sites in the plane. The first
    term in this Hamiltonian describes the coherent coupling of the ground and excited states with
     Rabi frequency $\Omega$, where $\hat \sigma^{(\vec i)}_{ge} = \ket{e_{\vec i}}\bra{g_{\vec
        i}}$ and $\hat \sigma^{(\vec i)}_{eg} = \ket{g_{\vec i}}\bra{e_{\vec i}}$ are the local
    transition operators. The second term is the van der Waals interaction
  potential between two atoms in the Rydberg state. In our case it is repulsive and takes the
  asymptotic form: $V_{\vec i\vec j} = -C_{6}/r_{\vec i\vec j}^{6}$, with the van der Waals
  coefficient $C_{6}<0$ and $r_{\vec i\vec j} = \alat |\vec i - \vec j|$ the distance between the
  two atoms at sites $\vec i$ and $\vec j$. The projection operator $\hat \sigma^{(\vec i)}_{ee} =
  \ket{e_{\vec i}}\bra{e_{\vec i}}$ measures the population of the Rydberg state at site $\vec
  i$. This model is valid as long as the mechanical motion of the atoms and all decoherence effects
  can be neglected (Supplementary Information).
  
  The dynamics of this strongly correlated system can be understood intuitively from its energy
  spectrum in the absence of optical driving. It is instructive to group the large number of
  many-body states, $2^\Nat$, according to the number of Rydberg excitations, $\Ne$, contained in
  each state (Fig.~\ref{fig:1}a). All singly excited states ($\Ne=1$) with different positions of
  the Rydberg atom have identical energies and form a \mbox{$\Nat$-fold} degenerate manifold. For
  multiply excited states ($\Ne > 1$), this degeneracy is lifted by the strong van der Waals
  interaction, giving rise to a broad energy band (Fig.~\ref{fig:1}a). Starting from the ground
  state, the creation of the first excitation is resonant, while the sequential coupling to
  many-body states with larger number of excitations is rapidly detuned by the interactions. In
  fact, the rapid variation of the van der Waals potential with distance prevents the excitation of
  all those states where Rydberg atoms are separated by less than the blockade radius, $\Rbk$,
  defined by $\hbar\Omega = -C_{6}/\Rbk^{6}$. The existence of this exclusion
  radius is expected to have a striking consequence: while the total many-body state exhibits finite-range correlations on a scale of $\Rbk$ \mycite{Robicheaux2005}, its high-density components with a Rydberg density close to $1/\Rbk^{2}$ should display a crystalline structure,
  meaning that the position of the Rydberg atoms is correlated over a distance comparable to the
  system size.

  The excitation dynamics of all configurations should occur in an
  entirely coherent fashion, resulting in highly non-classical many-body states. First, the
  approximate rotational symmetry of our system leads to symmetric superpositions of all microscopic configurations with different orientation but identical relative positions of the Rydberg
  atoms. Second, since the coupling addresses all states within an energy range
  $\sim\hbar\Omega$, it produces a coherent superposition of many-body states with different number of excitations and slightly different separation between the Rydberg atoms (Fig.~\ref{fig:1}a). This collective
  nature of the excited many-body states dramatically changes the timescale on which their dynamics
  occurs. The coupling strength to the state with a single excitation is enhanced by a factor
  $\sqrt{\Nat} \gg 1$ \mycite{Lukin2001} and the coupling to states with $\Ne > 1$
   is similarly enhanced, with $\Nat$ replaced by the number of energetically accessible
  configurations in each $\Ne$-manifold \mycite{Pohl2010a}.

  

  Our experiments began with the preparation of a two-dimensional degenerate gas of
  \numrange{150}{390} \Rb atoms confined to a single antinode of a vertical ($z$-axis) optical
  lattice \mycite{Endres:2011}. The gas was brought deep into the Mott-insulating phase by
  adiabatically turning on a square optical lattice with period $\alat = \SI{532}{nm}$ in the
  $xy$-plane. Within the system radius, $R =$ \SIrange{3.5}{5}{\micro m}, the probability of a
  lattice site to be occupied by a single atom was typically \SI{80}{\percent}. The atoms were then
  initialised in the hyperfine ground state $\ket g \equiv \ket{5S_{1/2}, F=2, m_{F}=-2}$ and
  coupled to the Rydberg state $\ket e \equiv \ket{43S_{1/2}, m_{J}=-1/2}$ (Fig.~\ref{fig:1}b). The
  coupling was achieved through a two-photon process via the intermediate state $\ket{5P_{3/2}, F=3,
    m_{F}=-3}$ using lasers of wavelengths \SIlist{780 ; 480}{nm} and $\sigma^{-}$ and $\sigma^{+}$
  polarisation, respectively (Fig.~\ref{fig:1}b and Methods). The resulting two-photon Rabi
  frequency was $\Omega/(2\pi) = \SI{170 +- 20}{kHz}$, yielding a blockade radius of $\Rbk = \SI{4.9
    +- 0.1}{\micro m}$. Following the initial preparation, we suddenly switched on the excitation
  lasers and let the system evolve for a variable duration $t$. After the excitation pulse, we
  detected the Rydberg excitations by first removing all atoms in the ground state with a resonant
  laser pulse, then deexciting the Rydberg atoms to the ground state via stimulated
  emission towards the intermediate state (Fig.~\ref{fig:1}b and Methods) and finally recording
    their position using high-resolution fluorescence imaging \mycite{Endres:2011}. The accuracy of the measurement
  was limited by the probability of \SI{75 +- 10}{\percent} to detect a Rydberg atom and by a
  background signal due to on average \num{0.2 +- 0.1} non-removed ground state atoms per picture
  (Supplementary Information). The spatial resolution of our detection technique is limited to
    about one lattice site by the residual motion of the atoms in the Rydberg state before
    deexcitation (Supplementary Information). Repeating the experiment many times allowed for
  sampling the different spatial configurations of Rydberg atoms constituting the
  many-body state and to measure their respective statistical weight.
    
  

  \begin{figure}
    \centering
    \includegraphics{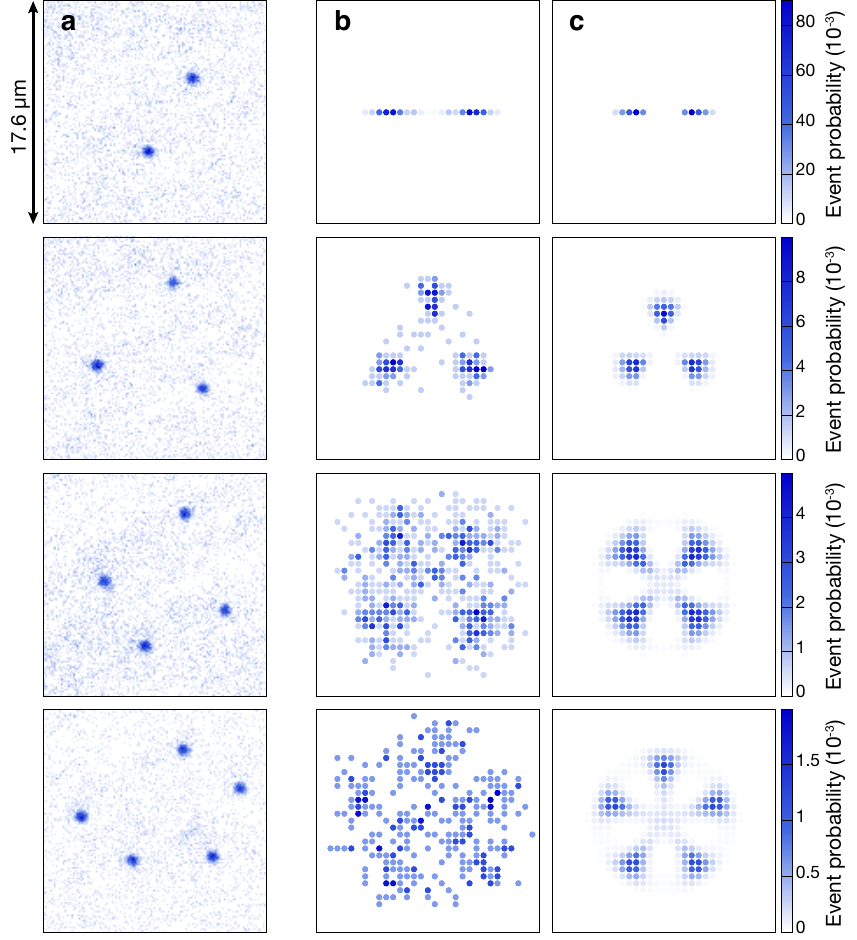}
    \caption{ {\bf Mesoscopic crystalline components of the many-body states.} Spatial distribution of
      excitations for the observed microscopic configurations sorted according to their number of
      excitations $\Ne = 2$-5 (top to bottom row). \textbf{a}, Examples of false-colour fluorescence
      images in which deexcited Rydberg atoms are directly visible as dark-blue spots. \textbf{b},
      Histograms of the spatial distribution of Rydberg atoms obtained after centring and aligning
      the individual microscopic configurations to a reference axis (Methods). The initial atom
      distribution had a diameter of $\SI{7.2+-0.8}{\micro\meter}$ and
      $\SI{10.8+-0.8}{\micro\meter}$ for $\Ne=2$-3 and $\Ne=4$-5, respectively. \textbf{c},
      Theoretical prediction from numerical simulations of the excitation dynamics governed by the
      many-body Hamiltonian of Eq.~\eqref{eq:H} for the same conditions as in the experiment
      (Supplementary Information).}
    \label{fig:2}
  \end{figure}


  In Fig.~\ref{fig:2}a we show typical images of microscopic configurations with
  \mbox{$\Ne=2$-5}. In order to analyse the structure of the many-body state, we group the individual images according to their
  number of excitations and determine the spatial distributions of the excitations, $\rho_e(\vec i)
  = \avg{\hat{\sigma}_{ee}^{(\vec i)}}$, where $\avg{\,\cdot\,}$ denotes the average from repeated
  measurements. These distributions display a typical ring-shaped profile (Fig.~\ref{fig:S1}),
  which results from the blockade effect and from the rotational symmetry of the system. Crystalline structures become visible once each microscopic configuration has been centred and
  aligned to a fixed reference axis (Fig.~\ref{fig:2}b and Methods). 
  
  For our smallest sample ($R\approx\SI{3.5}{\micro m}$), we observe 
  strong correlations between $\Ne=2$ excitations that are localized at a distance $\sim
  \SI{6}{\micro m}$, due to the interaction blockade. In the same dataset, configurations with \mbox{$\Ne = 3$} show an arrangement on an equilateral triangle, revealing both strong radial and azimuthal ordering. These
  correlations persist for larger numbers of Rydberg excitations, which we can prepare in larger samples ($R\approx\SI{5}{\micro m}$). They form quadratic and
  pentagonal configurations for $\Ne = 4$ and $\Ne = 5$, respectively. However, since their interaction energy is larger, these states are populated only with low probability, leading to a reduced
  signal-to-noise ratio. Our experimental data is in good agreement with numerical simulations of
  the many-body dynamics according to the Hamiltonian of Eq.~\eqref{eq:H}, for the same atom
  numbers, temperature and laser parameters as in the experiment (Fig.~\ref{fig:2}c and
  Supplementary Information). These simulations are based on a truncation of the underlying Hilbert space, exploiting the dipole blockade, and
    neglect any dissipative effects (\mycite{Pohl2010a} and Supplementary Information). The spatial distributions of excitations
  provided by the simulation reproduce all the features observed in the experiment. The only
  apparent discrepancy is the overall slightly larger size of the measured structures, which can
  be attributed to the spatial resolution of our detection method, as discussed below.

  

  \begin{figure*}[t]
    \centering
    \includegraphics{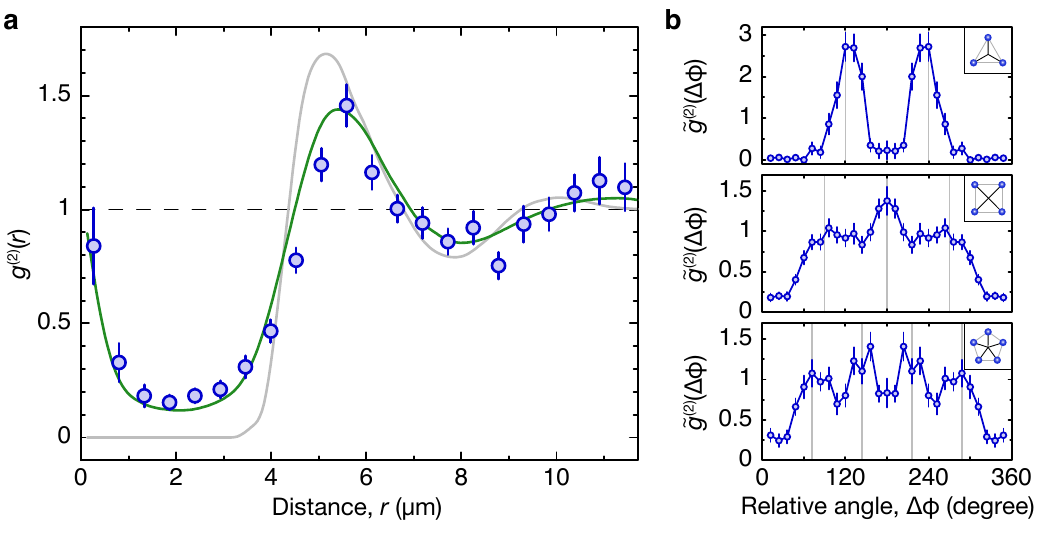}
    \caption{{\bf Correlation functions of Rydberg excitations.} \textbf{a}, Pair correlation
        function.  The blockade effect results in a strong suppression of the probability to find
        two excitations separated by a distance less than the blockade radius $\Rbk = \SI{4.9 +- 0.1}{\micro m}$. Moreover, we observe a peak at $r \simeq \SI{5.6}{\micro m}$ and a weak oscillation at larger distances. The initial atom
        distribution had a diameter of $\SI{10.8+-0.8}{\micro\meter}$. The experimental data (blue
        circles) are compared to the theoretical prediction both taking into account the independently characterized
        imperfections of our detection method (green line) and disregarding these imperfections
        (gray line). The dashed line marks the value of \gtwo in the absence of correlations. The
        error bars represent the standard error of the mean (s.e.m.) of $\gtwo(r)$.  \textbf{b},
        Azimuthal correlation function. The crystalline structure of the high-density components is best visible in the angular
        correlations around the centre of mass of the distribution of excitations, characterized by
        the correlation function $\tilde g^{(2)}(\Delta\phi)$ defined in Eq.~\eqref{eq:g2phi}.  By
        construction, this function is symmetric around \SI{180}{\degree}. Correlations are observed
        at the angles expected for the respective crystals shown in the insets. The peaks close to
        \SI{180}{\degree} are more pronounced since the centre of mass of a configuration is likely to lie close to the intersections of the diagonals, due to the
        blockade effect. Error bars, s.e.m.}
    \label{fig:3}
  \end{figure*}
	

  For a more quantitative analysis of 
  spatial correlations, we also measured the
  pair correlation function (Fig.~\ref{fig:3}a)
  \begin{equation}
    \label{eq:g2r}
    \gtwo(r) = \frac{\sum_{{\vec i} \neq {\vec j}} \delta_{r, r_{\vec i\vec j}} \, \langle
      \hat\sigma_{ee}^{(\vec i)} \hat\sigma_{ee}^{(\vec j)} \rangle} {\sum_{{\vec i} \neq {\vec j}}
      \delta_{r, r_{\vec i\vec j}} \, \langle \hat\sigma_{ee}^{(\vec i)} \rangle \langle
      \hat\sigma_{ee}^{(\vec j)} \rangle} \; ,
  \end{equation}
  which characterizes the occurrence of two excitations being separated by a distance $r$. Here
  $\delta_{r, r_{\vec i\vec j}}$ is the Kronecker symbol that restricts the sum to sites $(\vec i,
  \vec j)$ for which $r_{\vec i \vec j} = r$. In contrast to the spatial distributions presented
  above, the average is now taken over all values of $\Ne$. The pair correlation function $\gtwo(r)$
  shows a strong suppression at distances smaller than $r = \SI{4.8+-0.2}{\micro m}$, which
  coincides with the expected blockade radius $\Rbk=\SI{4.9+-0.1}{\micro\meter}$. Moreover, we find a clear peak at $r = \SI{5.6+-0.2}{\micro m}$ and
  evidence for weak oscillations extending to the boundaries of our system. This indicates that the overall many-body state only exhibits finite-range correlations. Our theoretical
  calculation of $\gtwo(r)$ (gray line in Fig.~\ref{fig:3}a) exhibits similar features, but shows
  more pronounced oscillations and vanishes
  perfectly within the blockade radius. These discrepancies can be attributed to several
  imperfections of the detection technique. The sharp peak at short distances $r \lesssim
  \SI{1}{\micro m}$ results from hopping of single atoms to adjacent sites during fluorescence
  imaging with a small probability of approximately \SI{1}{\percent}, which is falsely detected as
  two neighbouring excitations. The non-zero value of $\gtwo(r)$ for distances $r \lesssim
  \SI{3}{\micro m}$ arises from the imperfect removal of the ground state atoms. Finally, the
    shift and slight broadening of the peak in the correlation function is attributed to the
  residual motion of the Rydberg atoms before imaging (Supplementary Information). When accounting
  for these independently characterized effects in the theoretical calculations (green line in
  Fig.~\ref{fig:3}a), we recover excellent agreement with the measurements.
  
  
  Since our system size is comparable to the blockade radius, the excitations in states with $\Ne >
  1$ are localised along the
  circumference of the system. We characterize the resulting angular order by introducing an azimuthal correlation
  function that reflects the probability to find two excitations with a relative angle $\Delta\phi$
  measured with respect to the centre of mass of the distribution of excitations:
  \begin{equation}
    \label{eq:g2phi}
    \tilde g^{(2)}(\Delta\phi) = \int \frac{\text d\phi}{2\pi} \; \frac{\avg{\hat n(\phi) \hat
        n(\phi+\Delta\phi)}}{\avg{\hat n(\phi)} \avg{\hat n(\phi + \Delta\phi)}} \; .
  \end{equation}
  Here $\hat n(\phi) = \sum_{\vec i} \delta_{\phi, \phi_{i}} \, \hat \sigma_{ee}^{(\vec i)}$ is the
  azimuthal distribution of excitations, with $(r_{\vec i}, \phi_{\vec i})$ the polar coordinates of
  the site $\vec i$. As can be seen in Fig.~\ref{fig:3}b, the spatially ordered structure is clearly
  visible as correlations at relative angles $\Delta\phi = \nu \times \SI{360}{\degree}/\Ne$, with
  $\nu =1, 2, \ldots, \Ne$, even for the largest excitation numbers.
  
  

  \begin{figure*}
    \centering 
     \includegraphics[width=\textwidth]{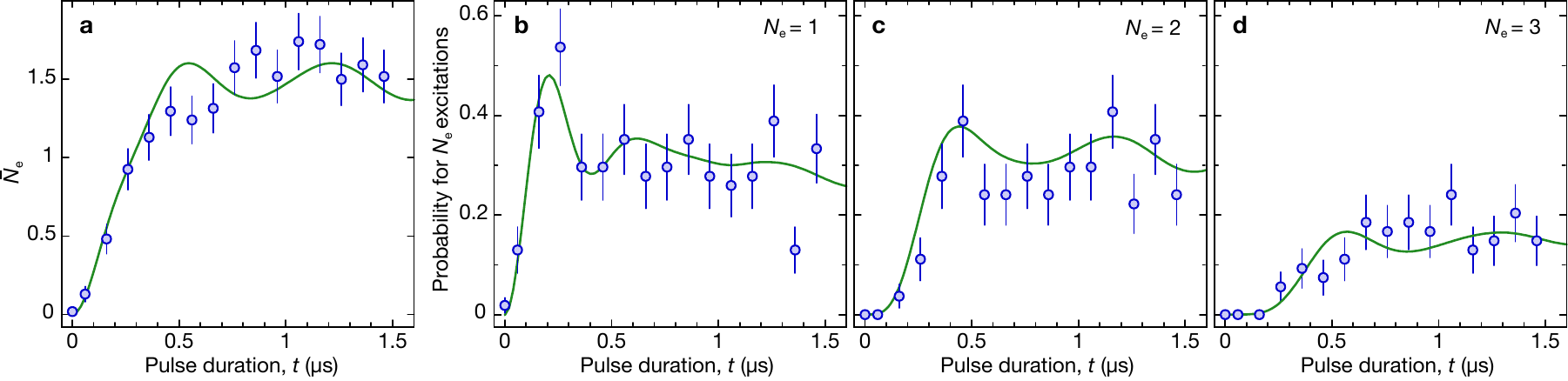}
    \caption{{\bf Time evolution of the number of Rydberg excitations.} \textbf{a}, Average number
      of detected Rydberg atoms as a function of the excitation pulse duration. Error bars,
      s.e.m. \textbf{b-d}, Time evolution of the probability to observe $\Ne = 1$ (\textbf{b}), $\Ne
      = 2$ (\textbf{c}) and $\Ne = 3$ (\textbf{d}) Rydberg excitations. The experimental data (blue
      circles) are compared to the theoretical prediction (green line), which is based on initial ground state atom distributions observed in
        the experiment and neglects all decoherence effects. It takes into account the finite
        detection efficiency as a free parameter (\SI{75}{\%}). Error bars, s.e.m.}
    \label{fig:4}
  \end{figure*}
  
   
  We finally analyse the many-body excitation dynamics of the system. In Fig.~\ref{fig:4}a we
   show the time evolution of the average number of Rydberg excitations, $\bar{\Ne} =
  \sum_{\vec i} \avg{\hat\sigma_{ee}^{(\vec i)}}$, which quickly saturates to a small value
  $\bar{\Ne} \simeq 1.5$, much smaller than the total number of atoms in the system, $\Nat =
  \num{150 +- 30}$. The saturation is reached in \SI{\sim 500}{ns}, 
  a factor of ten faster than the Rabi period $2\pi/\Omega$, due to the collective enhancement of
  the optical coupling strength.  The probability to observe $\Ne$ Rydberg excitations shows a
    similar saturation profile for each excitation number $\Ne$ (Fig.~\ref{fig:4}b-d), but on a
  timescale that increases with $\Ne$, from about \SI{200}{ns} for $\Ne=1$ to about \SI{600}{ns} for
  $\Ne=3$. This can be attributed to the variation of the collective
  enhancement factor associated with the number of energetically accessible microscopic configurations
 for a given $\Ne$. The theoretical
  excitation dynamics corresponding to the Hamiltonian \eqref{eq:H} shows remarkable agreement with
  the experimental data when including the finite detection efficiency. This provides evidence that
  the dynamics observed in the experiment is coherent, as expected on these timescales, which are
  much shorter than the lifetime of the Rydberg state of $\SI{25+-5}{\micro s}$ in the lattice
  and the timescale of other decoherence effects (Supplementary Information). The absence of
  high-contrast Rabi oscillations in the time evolution of the average number of Rydberg excitations
  is caused by the strong dephasing between many-body states with different interaction energies
  arising from the different spatial distribution of excitations. However, remnant signatures
  of Rabi oscillations can still be observed. In
  particular, the population of the singly excited states shows a peak around $t = \SI{200 +-
    50}{ns}$ (Fig.~\ref{fig:4}b), which matches the $\pi$-pulse time of the enhanced Rabi
  frequency $\pi/\left(\sqrt{\Nat}\Omega\right) = \SI{240 +- 40}{ns}$. Further evidence for the
  coherence of the dynamics can be found in the spatially resolved analysis of the excitation
  dynamics (Supplementary Information).


  
    In conclusion, we have characterised the strongly correlated excitation dynamics of a resonantly driven Rydberg gas using  optical detection 
    with unprecedented spatial resolution, and observed mesoscopic Rydberg crystals in the high-density components of the produced many-body states. One future challenge lies in the deterministic preparation of ground-state Rydberg crystals with a well-defined number of excitations via adiabatic sweeps of the laser parameters \mycite{Pohl2010a, Schachenmayer2010, VanBijnen2011}. Together with the demonstrated imaging technique, this would enable precise studies of quantum phase transitions in long-range interacting quantum systems on the microscopic level \mycite{Weimer:2008, Pohl2010a, Schachenmayer2010, Lesanovsky2011}. Combining the dipole blockade effect with the single-atom addressing demonstrated already in our experimental setup, one could also engineer mesoscopic quantum gates \mycite{Muller:2009}, which can serve as an experimental ``toolbox'' for digital quantum simulations of a broad class of spin models, including such fundamental systems as Kitaev's toric code \mycite{Weimer2010}.
  
    \section*{Acknowledgements}
  We thank R. L\"ow for discussions. We acknowledge funding by MPG, DFG, EU
  (NAMEQUAM, AQUTE, Marie Curie Fellowship to M.C.) and JSPS (Postdoctoral Fellowship for Research
  Abroad to T.F.).


  \bibliography{Rydberg_arxiv_final}



  \section*{Methods}
  
  \textbf{Rydberg excitation and detection scheme}  The two excitation laser beams were counterpropagating along the $z$-axis, with an
  intermediate-state detuning \linebreak $\delta/(2\pi)=\SI{742+-2}{MHz}$
  (Fig.~\ref{fig:1}b). During the sequence, a magnetic offset field of \mbox{$B \simeq \SI{30}{G}$}
  along the $z$-axis defined the quantisation axis. The excitation pulse was performed by switching
  the laser at \SI{780}{nm} while the laser at \SI{480}{nm} was on. The temporal resolution of our
  measurement was thus set by the rise time of the \SI{780}{nm} light, which was $\simeq
  \SI{40}{ns}$. Immediately after the excitation pulse, we used near-resonant circularly-polarised
  laser beams to drive the transitions $\mbox{\ket{5S_{1/2}, F=1}} \rightarrow \mbox{\ket{5P_{3/2},
      F=2}}$ and $\mbox{\ket{5S_{1/2}, F=2}} \rightarrow \mbox{\ket{5P_{3/2}, F=3}}$ and remove all
  ground state atoms, with a fidelity of \SI{99.9}{\%} in \SI{10}{\micro s}. Subsequently, the
  Rydberg atoms were stimulated down to the ground state by resonantly driving the
  $\mbox{\ket{43S_{1/2}, m_{J}=-1/2}} \rightarrow \mbox{\ket{5P_{3/2}, F=3, m_F=-3}}$ transition for
  \SI{2}{\micro\second}.
  
  \textbf{Computation of the histograms}  The histograms shown in Fig.~\ref{fig:2}b are based on the digitised atom distribution
  reconstructed from the raw images \mycite{Endres:2011}. They reflect the Rydberg atom
  distribution in a region of interest covering a disc of radius $R_{\text{max}} = 1.5 \times
  R$. Each individual image was aligned in the following way. First, we set the origin of the
  coordinate system to the centre of mass of the atom distribution. Then, for each atom we
  determined the angle between its position vector and a reference axis, and rotated the images
  about the origin by the mean value of these angles (repeating this operation would leave the
  configuration unchanged). The histograms contain data taken at different evolution times up to
  \SI{4}{\micro\second}, as we found no significant temporal dependence of the excitation patterns. The theoretical calculations used the same parameters as in the experiment (including
  temperature and atom number distribution of the initial state) and followed the same procedure to
  determine the Rydberg atom densities. Both the experimental and theoretical histograms were
  normalised such that the value at each bin represents the probability to observe a microscopic
  configuration with a Rydberg atom located at this position.
 
\section*{Supplementary Information}
  
  \section{Spatial distribution of the excitations without rotational alignment}

  Here we show the spatial distribution of excitations based on the same data as in Fig.~2b and 2c
  of the main text but without the rotational alignment procedure described in the Methods section.
  \vspace{1cm}
  \begin{figure}[b]
    \centering
    \includegraphics{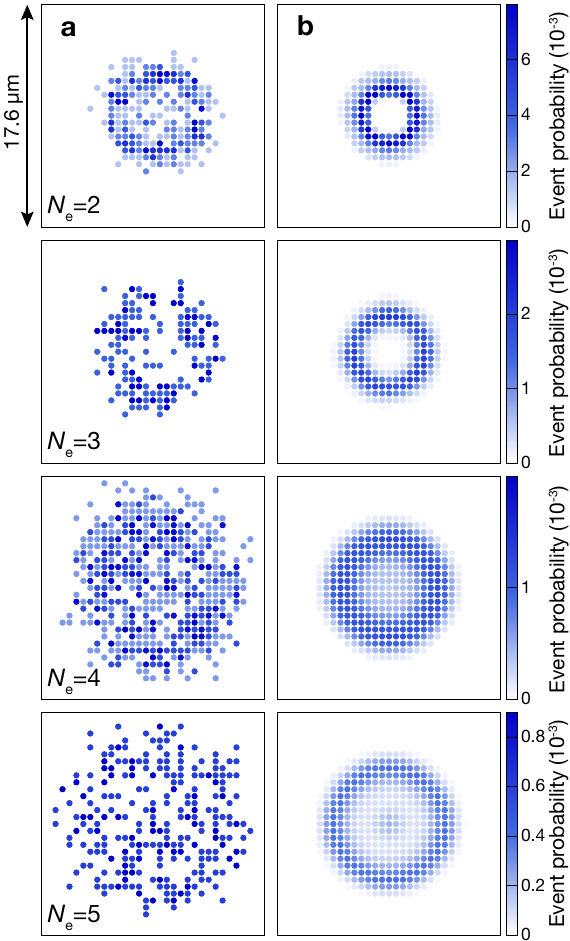}
    \caption{{\bf Spatial distribution of the distribution of excitations before rotation.}
      \textbf{a}, Histograms constructed from the experimental data. The Rydberg atoms are excited
      with higher probability close to the edge than in the centre of the cloud. The resulting
      ring-shaped excitation region is clearly visible for $\Ne=2$ and 3. The contrast decreases for
      $\Ne=4$ and 5 due to the lower number of occurences in the experiment. \textbf{b}, Theoretical
      predictions for the excitation from initial clouds of same temperature and atom number as in the
      experiment (see Fig.~2 for details).}
    \label{fig:S1}
  \end{figure}
  
  
  \section{Spatially resolved analysis of the excitation dynamics}
  
\begin{figure}[t]
    \centering
    \includegraphics[width=\columnwidth]{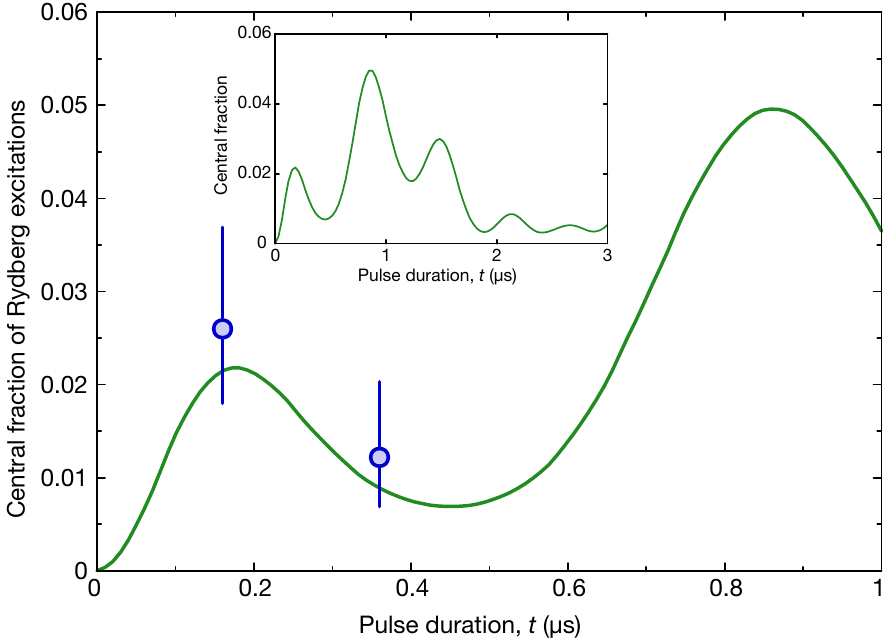}
    \caption{{\bf Excitation dynamics at the centre of the system.} Relative number of excitations
      in the central nine sites as a function of the excitation pulse duration for microscopic
      configurations with a single excitation $\Ne=1$. The theoretical calculation (green line,
      inset) reveals the coherent evolution, which is hardly visible in the time evolution of the
      total excitation number. Two experimental points (blue circles) were obtained from an
      additional dataset containing about 800 images per pulse duration. It was characterized by the
      temperature of the initial state $T = \SI{9+-2}{nK}$, the atom number $\Nat = \num{210+-30}$
      and the radius $R = \SI{4.2+-0.5}{\micro m}$. These experimental parameters were included in
      the numerical simulation. The error bars denote one standard deviation of the mean (s.e.m.).}
    \label{fig:S2}
  \end{figure}

  The coherence of the dynamics can be revealed in a more obvious way by studying the time evolution
  of the spatial distribution of the Rydberg excitations. For this purpose, we considered the subset
  of microscopic configurations with only one excitation. Because the blockade radius is only
  slightly smaller than the system diameter, only those configurations in which the excitation is
  located close to the edge of the system are significantly coupled to configurations with two
  excitations. This results in different time constants for the dynamics at different distance $r$
  from the centre. We have investigated this effect theoretically by calculating the time evolution of
  the relative probability for the excitation to be located close to the centre of the system (green
  line in Fig.~\ref{fig:S2}). In contrast to Fig.~4 of the main text, we now observe Rabi-like
  oscillations with notable amplitude over long timescales. We performed the corresponding
  measurement in the experiment for two pulse durations (blue circles) and find reasonable
  agreement. This provides further evidence for the coherence of the many-body dynamics in the
  experiment.
  
  
  \section{Additional information on the datasets}

  The experimental data results from three different datasets (A, B and C). Each dataset was
  characterized by a temperature, $T$, atom number, $\Nat$, and diameter, $2R$, which we extracted from
  a fit to the ground state atom distribution in the initial state \mycite{Sherson:2010}
  (Table~\ref{tab:1}). The datasets A and B were used for Fig.~2 and 3, while the dataset C was
  used for Fig.~4. The distribution of the number of excitations in the datasets A and B is detailed
  in Table~\ref{tab:2}, where we also indicated which subset of images was used for which
  figures. The dataset C consisted of 54 images per pulse duration and the relative distribution of
  excitations is directly visible in Fig.~4.
  \vspace{1cm}
   \newcommand{\minitab}[2][l]{\begin{tabular}{#1}\vspace{-1pt}#2\end{tabular}}
   \begin{table}[h]
    \centering
    \begin{tabular}{lrrr}
      \toprule & \multicolumn{1}{c}{dataset A} & \multicolumn{1}{c}{dataset B} & \multicolumn{1}{c}{dataset C} \\
      \midrule
      T (\si{nK}) & \tablenum[table-format=3]{8 +- 4}  & \tablenum[table-format=3]{13 +- 2} & \tablenum[table-format=3]{9 +- 4} \\
      $\Nat$ & \tablenum[table-format=3]{150 +- 30} & \tablenum[table-format=3]{390 +- 30} &  \tablenum[table-format=3]{150 +- 30} \\
      R (\si{\micro m}) & \tablenum[table-format=3]{3.6+-0.4} & \tablenum[table-format=3]{5.4 +- 0.4} & \tablenum[table-format=3]{3.6+-0.4} \\
      \bottomrule
    \end{tabular}
    \caption{Temperature $T$, atom number $\Nat$ and radius $R$ for the datasets A, B and C. Errors, s.d.}
    \label{tab:1}
  \end{table}
  \vspace{1cm}
  \begin{table}
    \centering
    \begin{tabular}{ccccc}
      \toprule & \multicolumn{2}{c}{dataset A} & \multicolumn{2}{c}{dataset B} \\
      \cmidrule(rl){2-3} \cmidrule(rl){4-5}
      \multirow{2}*{\minitab[c]{number of \\ excitations}} & \multirow{2}*{\minitab[c]{number of \\ images}} &
      \multirow{2}*{\minitab[c]{figures}}  & \multirow{2}*{\minitab[c]{number of \\ images}} &
      \multirow{2}*{\minitab[c]{figures}}  \\ \\
      \midrule
      0 & \tablenum[table-format=3]{177}  & -- & \tablenum[table-format=3]{321} & -- \\
      1 & \tablenum[table-format=3]{235} & -- & \tablenum[table-format=3]{375} & 3a \\
      2 & \tablenum[table-format=3]{191} & 2a, 3b & \tablenum[table-format=3]{390} & 3a \\
      3 & \tablenum[table-format=3]{65} & 2a, 3b & \tablenum[table-format=3]{308} & 3a \\
      4 & \tablenum[table-format=3]{7} & -- & \tablenum[table-format=3]{177} & 2a, 3a, 3b \\
      5 & \tablenum[table-format=3]{1} & -- & \tablenum[table-format=3]{64} & 2a, 3a, 3b \\
      6 & \tablenum[table-format=3]{0} & -- & \tablenum[table-format=3]{14} & -- \\
      7 & \tablenum[table-format=3]{0} & -- & \tablenum[table-format=3]{5} & -- \\
      \bottomrule
    \end{tabular}
    \caption{Distribution of the number of excitations in the datasets A and B. For each subset of
      images we have also indicated in which figure of the main text it has been used.}
    \label{tab:2}
  \end{table}

  
  \section{Numerical calculations}

  \begin{figure*}
    \centering
    \includegraphics[width=\textwidth]{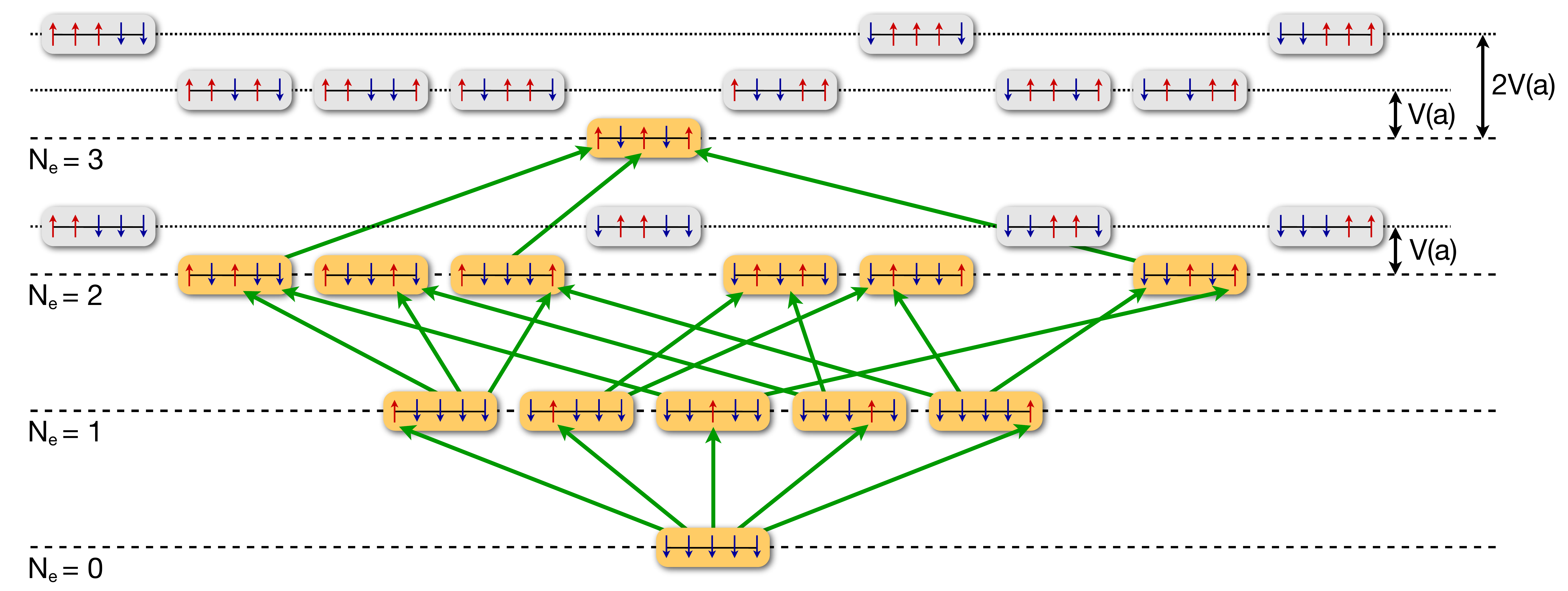}
    \caption{{\bf Schematics of the numerical calculations.}  The underlying many-body level
      structure is shown for the example of a one-dimensional chain of five atoms. The
      atomic states are symbolised by effective spins, with spin-down (blue arrows) and spin-up (red
      arrows) corresponding to the atomic ground and Rydberg state, respectively. In the displayed
      example, we consider strong interactions $V(a) \gg \hbar\Omega$ between adjacent Rydberg
      excitations, while next-nearest neighbour interactions $V(2a) = V(a)/64 < \hbar\Omega$ are
      assumed to be smaller than the laser coupling strength. Consequently, the dynamics of near
      resonant basis states (orange boxes) is explicitly calculated in the simulations, while
      strongly shifted states (grey boxes) do not participate in the excitation dynamics and are
      discarded (see text for further details). The near-resonant laser coupling between relevant
      many-body states is indicated by the green arrows. Due to the strong geometrical constraint
      imposed by the interaction blockade combined with the finite system size, many-body states
      containing more than $\Ne=3$ Rydberg excitations do not need to be considered.}
    \label{fig:S3}
  \end{figure*}  
  In order to determine the dynamics governed by the Hamiltonian in Eq.~(1), we expand the many-body
  wave function, $|\psi\rangle$, of the \Nat-atom system in terms of Fock-states
  \begin{multline*}
    |\psi\rangle = c^{(0)}|0\rangle + \sum_{{\vec i}_1} c^{(1)}_{{\vec i}_1}|{\vec i}_1\rangle +
    \sum_{{\vec i}_1, {\vec i}_2} c^{(2)}_{{\vec i}_1,{\vec i}_2}|{\vec i}_1, {\vec i}_2\rangle +
    \ldots \\ + \sum_{\mathclap{{\vec i}_1, \ldots, {\vec i}_{\Nat}}} c^{(\Nat)}_{{\vec i}_1, \ldots,
      {\vec i}_{N_{\rm at}}}|{\vec i}_1, \ldots, {\vec i}_{\Nat}\rangle \; ,
	\end{multline*}
  where $|{\vec i}_1, \ldots, {\vec i}_{N_{\rm e}}\rangle$ corresponds to a state with $N_e$ Rydberg
  excitations located at lattice sites ${\vec i}_1$ to ${\vec i}_{\Ne}$, and $c^{(N_{\rm e})}_{{\vec
      i}_1, \ldots, {\vec i}_{N_{\rm e}}}$ denotes the respective time dependent amplitude. The
  basis states are eigenfunctions of the Hamiltonian (1) in the absence of laser driving, with
  energy eigenvalues $E^{(\Ne)}_{{\vec i}_1,...,{\vec i}_{\Ne}} = \sum_{\alpha<\beta}^{\Ne} V_{{\vec
      i}_\alpha{\vec i}_\beta}$. For a system of \Nat atoms, this basis set expansion yields a set
  of $2^{\Nat}$ coupled differential equations (Fig.~\ref{fig:S3}). Due to the exponential
  growth of the number of many-body states with \Nat, a direct numerical propagation is practically
  impossible for the large number of atoms in our experiments, $\Nat \sim 100$. In order to make the
  calculations feasible, we exploit the blockade effect and discard all many-body states containing
  Rydberg atom pairs separated by less than a critical distance $R_{\rm c}$. For the present
  simulations, we obtain well-converged results for $R_{\rm c} \simeq \Rbk/2$, where \Rbk is the
  blockade radius. The resulting geometric constraint not only reduces the number of relevant
  many-body states within a given \Ne-manifold, but, due to the finite system size, also restricts
  the total number of excitations \Ne necessary to obtain converged results. For the parameters
  considered in this work, a maximum number of Rydberg excitations of $\Ne^{({\rm max})}=6$ was
  found sufficient. This procedure allows to significantly mitigate the otherwise strong exponential
  scaling of the underlying Hilbert space dimension, and yields a power-law dependence $\sim N_{\rm
    at}^{N_{\rm e}^{({\rm max})}}$ of the number of relevant basis states on the total number of
  atoms. This makes the computations feasible, albeit still demanding, for such large systems as in
  our experiment.

  
  \section{Optical detection scheme}
  
We developed a fully-optical detection technique for Rydberg atoms, which represents
  an alternative to the usual detection schemes based on the ionisation of Rydberg atoms
  \mycite{Low2012, Schwarzkopf2011, Urban2009, Potvliege2006}. Here, we provide additional
  details to those given in the main text, especially regarding the detection efficiency and spatial
  resolution.

  \subsection{Stimulated deexcitation of the Rydberg atoms}
  
  We stimulated the Rydberg atoms down to the ground state by resonantly driving the
  $\ket{43S_{1/2}, m_{J}=-1/2} \rightarrow \ket{5P_{3/2}, F=3, m_F=-3}$ transition. The Rabi
  frequency associated with this resonant single-photon transition was typically several \si{MHz}.
  In combination with the short lifetime of the $5P_{3/2}$ state (\SI{27}{ns}), this
  allows for a very efficient and fast (\SI{<2}{\micro s}) pumping to the ground state. The laser light
  resonant with the transition between the Rydberg and the intermediate states was produced by a
  resonant electro-optical modulator (EOM) in the path of the excitation laser at \SI{480}{nm},
  which created a sideband at the desired intermediate-state detuning $\delta/(2\pi) =
  \SI{742}{MHz}$. The other sideband and the carrier have negligible influence on the atoms in the
  deexcitation phase since they are off-resonant. The EOM also allows for the required fast
  switching of the deexcitation light within \SI{<1}{\micro s}.
  
  \subsection{Spatial resolution}\label{sec:spatialresolution}
  
  Two effects can in principle limit the spatial resolution of our detection technique. The first
  one is the residual hopping of the atoms during the fluorescence imaging phase. We found that such
  an event can occur in our experiment with a probability of approximately \SI{1}{\percent} per
  particle. When this happens, the moving atom will yield a fluorescence signal on two adjacent
  sites, which can be falsely attributed to two distinct atoms by the reconstruction algorithm. This
  detection artifact results in a correlation signal at short distances $r < \SI{1}{\micro\meter}$
  (Fig.~3a).  However, due to their rarity these events have negligible influence on the spatial
  resolution. The second effect is the possible motion of the Rydberg atoms in the optical lattice
  potential before the imaging phase. For Rydberg atoms, the lattice potential has similar
  amplitude but opposite sign compared to ground state atoms
  \mycite{Anderson2011}. An excited atom therefore finds itself at a maximum
  of the periodic potential and can move in the $xy$-plane with an average velocity of
  $\sim\SI{30}{nm/\micro s}$, for a typical depth of the optical lattice potential of $V_\text{lat}
  = \SI{40}{\Er}$. Both effects lead to a possible motion of the Rydberg atoms by about
  one lattice site during the \SI{10}{\micro s} of the removal pulse. 
  The recoil velocity acquired in the two-photon excitation process is
  insignificant in comparison as it is oriented along $z$-direction and only of smaller magnitude
  $\simeq\SI{4}{nm/\micro s}$. 

  \subsection{Detection efficiency}
	
  The detection efficiency for Rydberg atoms in our setup is limited by the lifetime of the Rydberg
  state and by the anti-confining character of the optical lattice potential for the Rydberg
  atoms. If a Rydberg atom decays to the ground state during the removal pulse, it will be removed
  as well and not detected. The residual motion of the Rydberg atoms in the lattice potential
  also leads to a reduction of the detection efficiency when the atoms move away from the focal
  plane of the imaging system, which has a depth of focus of order \SI{1}{\micro m}. Both effects
  can be reduced by minimising the time the atoms spend in the Rydberg state after the excitation
  pulse, by reducing the duration of the removal pulse for the ground state atoms. A
  removal pulse duration of \SI{10}{\micro s} turned out to offer the best compromise between a good
  detection efficiency of Rydberg atoms and a low survival probability of ground state
  atoms.

  We estimated the detection efficiency in three different ways. First, we measured the lifetime of
  the atoms in the Rydberg state in the lattice by varying the duration of the removal pulse. We fitted a
  $1/e$-decay time $\tau = \SI{25+-5}{\micro s}$, which corresponds to a detection efficiency of
  \SI{65+-5}{\percent}. A second estimation is provided by the time evolution of the number of
  Rydberg excitation displayed in Fig.~4. Here the theoretical prediction matches the data best when
  assuming a detection efficiency of \SI{\sim 75}{\percent}, which is compatible with the previous
  estimate. Finally, the statistical weight of the different number of excitations can also be
  related to the detection efficiency. This last estimation points to a higher detection efficiency
  of \SI{\sim 80}{\percent}. Combining all these values with equal weight, we finally obtain a
  detection efficiency of \SI{75}{\percent} with an uncertainty of about \SI{10}{\percent}.
	
  \section{Validity of the the model}

  The validity of the Hamiltonian in Eq.~(1) for our experimental system relies on two main
  assumptions, which are discussed in this section: the positions of the atoms is frozen during the
  dynamics and all decoherence sources can be neglected.

  \subsection{Movement of the atoms during the dynamics}
  
  The ground state atoms were confined in a three-dimensional optical lattice of depth $V_{xy} =
  \SI{40 +- 3}{\Er}$ in the $xy$-plane and $V_{z} = \SI{75 +- 5}{\Er}$ along the $z$-axis, where
  $\Er = (2\pi\hbar)^{2}/(8 m a_{\text{lat}}^{2})$ denotes the recoil energy of the lattice, and $m$
  the atomic mass of \Rb. For the minimum lattice depth used in the experiment of $V_{\text{lat}} =
  \SI{40}{\Er}$, the time associated to the inverse of the tunnelling matrix element was
  \mbox{$\hbar/J \simeq \SI{700}{ms}$}, and therefore negligible compared to the timescale of the
  internal dynamics.  The Rydberg atoms move in the lattice potential with a typical velocity of
  \SI{30}{nm/\micro s} (see discussion in Section~\ref{sec:spatialresolution}), which can also be
  neglected.
  
  \subsection{Light scattering from the intermediate state}

  One source of decoherence in our experimental system is light scattering from the intermediate
  state used in the two-photon excitation process. The laser beam off-resonantly driving the
  ground-to-intermediate-state transition had a detuning of \SI{742}{MHz} and an intensity of
  \SI{\sim 450}{mW/cm^2}, 
  yielding a scattering rate of \SI{9e4}{s^{-1}}. This corresponds to a coherence time of
  \SI{11}{\micro s}, which is a factor of ten longer than the typical timescale of the
  many-body dynamics.
 
  \subsection{Laser linewidth}
 
  The finite spectral width of the optical radiation driving the transition to the Rydberg state
  acts as a decoherence source and was reduced by carefully stabilising the frequency of the
  excitation lasers. We could achieve a two-photon linewidth of \SI{\approx70}{kHz}, leading to a
  coherence time of \SI{14}{\micro\second}. Technical details on the laser setup are provided in
  Section~\ref{sec:lasersetup}.
 
  \subsection{Two-photon Rabi frequency}
 
  We determined the two-photon Rabi frequency by driving Rabi oscillation in a very dilute system,
  where the average distance between two atoms was larger than the blockade radius of the $43S$ state, for which the van der Waals coefficient is $C_{6} = \SI{-1.6e-60}{J.m^{6}}$ \mycite{Singer2005}. The measurement
  was very time-consuming in our experimental setup since at such densities only a few atoms are
  located within the waist of the excitation lasers, resulting in a very low signal (about one
  Rydberg excitation per image) with relatively large fluctuations. We could observe one period of
  the Rabi oscillation, as expected from the combined coherence time of light scattering and laser
  linewidth of \SI{6}{\micro\second}, and extracted from the data a Rabi frequency of $\Omega/(2\pi)
  = \SI{170+-20}{kHz}$.

  The waists of the two laser beams of wavelength \SI{780}{nm} and \SI{480}{nm}  \mycite{Mack2011} driving the
  two-photon transition were \SI{57+-2}{\micro m} and \SI{17+-5}{\micro m}, respectively. The
  largest systems we studied had a radius of \SI{5.4}{\micro m}, leading to a variation in the
  coupling strength to the Rydberg state by \SI{<20}{\percent} over the whole system.

  \vspace*{0.5cm}
  
  \section{Laser setup}\label{sec:lasersetup}
 
  The light at a wavelength of \SI{780}{nm} was produced by a diode laser whose frequency was
  stabilised using a modulation transfer spectroscopy in a rubidium vapour cell. The light at
  \SI{480}{\nano\meter} was generated by frequency-doubling light at \SI{960}{nm}, which was emitted
  from a diode laser and amplified by a tampered amplifier. This second laser was stabilised by a
  phase-lock to a master laser, which allows for tuning its frequency while maintaining the narrow
  laser linewidth. The master laser at \SI{960}{nm} was locked to a temperature stabilised ULE
  cavity in a vacuum chamber. The short-term linewidth of both excitation lasers was measured using
  an independent resonator (EagleEye, Sirah Laser- und Plasmatechnik GmbH, Germany), which can
  resolve linewidths down to \SI{\sim 20}{kHz}. We obtained a linewidth of \SI{20}{kHz} for the
  laser at \SI{480}{nm} and \SI{50}{kHz} for the laser at \SI{780}{nm}. We measured the long-term
  stability of the two-photon excitation to be \SI{50}{kHz} over several hours (FWHM of the centre
  of the line) using EIT-spectroscopy in a rubidium vapour cell \mycite{Mohapatra2007}.

\end{document}